\documentclass[useAMS,usenatbib]{mn2e}
\newcommand{\be}{\begin{equation}}    % for lazy typers
\newcommand{\ee}{\end{equation}}
\newcommand{\ba}{\begin{eqnarray}}
\newcommand{\ea}{\end{eqnarray}}

\title[Resonances and bifurcations in axisymmetric scale-free potentials]
{Resonances and bifurcations in axisymmetric scale-free potentials}
\author[Giuseppe Pucacco]{Giuseppe Pucacco$^{1,2}$\thanks{E-mail:
pucacco@roma2.infn.it}\\
$^{1}$Physics Department, University of Rome ``Tor Vergata", I00133 Rome\\
$^{2}$INFN, Sezione Roma Tor Vergata
}
\begin{document}

\date{Accepted 2009 June 15. Received 2009 May 20}

\pagerange{\pageref{firstpage}--\pageref{lastpage}} \pubyear{2009}

\maketitle

\label{firstpage}

\begin{abstract}
We investigate an analytical treatment of bifurcations of families of resonant `thin' tubes in axisymmetric galactic potentials. We verify that the most relevant bifurcations are due to the (1:1) resonance producing the `inclined' orbits through two different mechanisms: from the disk orbit and from the `thin' tube associated to the vertical oscillation. The closest resonances occurring after these are the (4:3) resonance in the oblate case and the (2:1) resonance in the prolate case. The (1:1) resonances are treated in a straightforward way using a 2nd-order truncated normal form. The higher-order resonances are instead cumbersome to investigate, because the normal form has to be truncated to a high degree and the number of terms grows very rapidly. We therefore adopt a further simplification giving analytic formulas for the values of the parameters at which bifurcations ensue and compare them with selected numerical results. Thanks to the asymptotic nature of the series involved, the predictions are reliable well beyond the convergence radius of the original series. 

\end{abstract}


\section{Introduction}

Axisymmetric galactic potentials admit only tube orbits around the symmetry axis \citep{bt}. However, departing from the two linear approximations around the circular orbit, epicyclic motion in the symmetry plane and the oscillation perpendicular to this plane, nonlinear dynamics are characterized by a hierarchy of resonances. To these are associated the bifurcations of different kinds of `thin' tubes which parent the corresponding `thick' families \citep{conto2}. The occurrence of this phenomenon may heavily affect the kinematics of stars with, for example, sudden removal from the disk ({\it levitation}) and/or velocity redistribution that can be viewed as a heating mechanism of the disk \citep{ST1}. Resonant families may also play a relevant role in self-consistent models of spheroidal halo or disk/halo systems.

Scale-free potentials are often adopted as realistic galactic models \citep{rich,eva,TT, ST2}. The orbit structure is the same at each energy level, resulting in a much simpler description of dynamics. Their axisymmetric version is parametrized by only two parameters: $\alpha$, the exponent of the power law (which can be taken zero in the limit case of the logarithmic potential) and $q$, the ellipticity of the isopotentials. Adding to these $L$, the conserved angular momentum around the symmetry axis, completely specifies the orbit structure. The ensuing 2-dof dynamical system is in general non-integrable. An analytic approach can be attempted by expanding the effective potential around the critical point corresponding to the circular orbit and pushing the expansion beyond quadratic terms. The Hamiltonian series is afterwards replaced by a `normal form' which plays the role of an integrable approximation of the original non-integrable system \citep{conto1,gu,gs,ya}. The dynamics of the normal form are amenable to a totally analytic treatment that allows us, among other things, to find the bifurcation thresholds of periodic orbits in terms of the parameters of the effective potential \citep{BBP1,BBP2} and approximate expressions of the solutions of the equations of motion \citep{PBB2}. 

In the present work, we investigate the properties of the simplest non-trivial series expansion of axisymmetric scale-free potentials beyond the epicyclic approximation. We verify that the most relevant bifurcations are due to the (1:1) resonance producing the `inclined' orbits through two different mechanisms \citep{hu}: from the disk orbit (the `horizontal' normal mode) and from the `thin' tube associated to the $z$ oscillation (the `vertical' normal mode) The closest resonances occurring after these are the (4:3) resonance in the oblate case and the (2:1) resonance in the prolate case. Other high-order resonances can be present at very low values of the angular momentum.

The (1:1) resonances are treated in a straightforward way using a 2nd-order truncated normal form. The higher-order resonances are instead cumbersome to investigate, because the normal form has to be truncated to a high degree and the number of terms grows very rapidly. We therefore adopt a further simplification suggested by the approach of \citet{SV}. We give therefore analytic formulas for the values of $\alpha, q, L$ at which bifurcations ensue and compare them with selected numerical results. Thanks to the asymptotic nature of the series involved \citep{PBB1}, the predictions are reliable well beyond the convergence radius of the original series. 

The plan of the paper is as follows: in section 2 we describe the family of systems under investigation and how they are prepared to undergo a perturbative analysis; in section 3 we recall the method based on Hamiltonian normal forms; in section 4 we apply the theory to compute bifurcation thresholds of the main resonant families; in section 5 we conclude with a discussion of the results.

\section{The potentials and their series expansions}

We investigate the dynamics of the family of potentials
\be\label{pota}
\Phi_{\alpha} (R,z;q) =\cases { & $ {{1} \over {|\alpha|}} \left(R^2 + \frac{z^2}{q^2}\right)^{\alpha/2}, \;\; \alpha \ne 0, $\cr
& $\frac12 \log\left(R^2 + \frac{z^2}{q^2}\right), \;\;  \alpha = 0.$\cr}\ee
The ellipticity of the equipotentials is determined by the parameter $q$: we have an `oblate' figure when $q < 1$ and a `prolate' figure when $q > 1$. The slope of the power law will be restricted to the range
\be\label{alfa}
-1 < \alpha \le 2. \ee
The Hamiltonian of the system in cilindrical coordinates is 
\be\label{Ha}
H = \frac12 \bigg(p_{R}^{2} + \frac{p_{\phi}^{2}}{{R}^{2}} + p_{z}^{2} \bigg) + \Phi_{\alpha},\ee
that, exploiting the conservation of the axial angular momentum
\be\label{am}
p_{\phi} = L,\ee
is effectively the Hamiltonian of a two degrees of freedom system in the family of potentials
\be\label{pote}
V_{\alpha} (R,z; L,q) = \frac{L^{2}}{2{R}^{2}} + \Phi_{\alpha} (R,z;q).\ee
These potentials have a unique absolute minimum in 
\be\label{amin}
R=R_{\rm c}(\alpha)=L^{\frac2{2+\alpha}},\;\;z=0.\ee
This is a stable equilibrium on the {\it meridional} plane $\phi = {\rm const}$ corresponding to a circular orbit of radius $R_{\rm c}(\alpha)$ of the full three-dimensional problem. Since the dynamics are scale-free we may fix energy once and for all
\be\label{ena}
E \doteq E_{\alpha} = \cases { &$ \left(\frac12 + \frac{1}{\alpha}\right){\rm e}^{-\frac{\alpha}{2+\alpha}}, \;\; \alpha \ne 0, $\cr
&$ 0, \;\;  \alpha = 0.$\cr}\ee
This implies that the radius of the circular orbit at this energy is
\be\label{circ}
R_{\rm c}(\alpha)={\rm e}^{-\frac{1}{2+\alpha}}, \;\; -1 < \alpha \le 2\ee
and we can investigate the dynamics at 
\be
E=E_{\alpha}, \;\; \forall \alpha \in (-1, 2],\ee 
by varying $L$ in the range
\be\label{amr}
0 < L \le L_{\rm max} \equiv R_{\rm c}^{\frac{2+\alpha}2} = \frac1{\sqrt{\rm e}}\ee
without any loss of generality \citep{hu}.

In order to implement the perturbation method we expand the effective potential around the minimum (\ref{amin}). A common attitude in normal form theory holds up the reliability of its predictions well beyond the convergence radius of the expansion. Although a rigorous proof of this statement is still lacking, it is confirmed by several results obtained in analytical and celestial mechanics \citep{scu,ECG,PBB1}. One of the aims of the present work is to test the reliability of these asymptotic estimates.

We introduce rescaled coordinates according to
\be\label{xy}
x \doteq \frac{R-R_{\rm c}}{R_{\rm c}}, \quad y \doteq \frac{z}{R_{\rm c}}\ee
with origin in the equilibrium point (\ref{amin}). The potential  (\ref{pote}) is expanded as a truncated series (in the coordinates $x,y$) of the form
\be\label{potes}
V_{\alpha}^{(N)} (x,y;L,q) = 
\sum_{k=0}^{N} \sum_{j=0}^{k} C_{(j,k-j)}(L,\alpha,q)x^{j}y^{k-j},
\ee
where the truncation order $N$ is determined by the resonance under study and is discussed below. From (\ref{amin}) and the rescaling (\ref{xy}), the constant term of the expansion is
\be\label{CZZ}
C_{(0,0)} (L,\alpha) =\cases { & $ \left(\frac12 + \frac{1}{\alpha}\right)L^{\frac{2\alpha}{2+\alpha}}, \;\; \alpha \ne 0, $\cr
& $\frac12 + L, \;\;  \alpha = 0$\cr}\ee
and the other coefficients have the form
\be\label{CJK}
C_{(j,k-j)} (L,\alpha,q) = L^{\frac{2\alpha}{2+\alpha}} c_{(j,k-j)} (\alpha,q).
\ee
In order to simplify formulas, we introduce the new parameter
\be\label{beta}
\beta = -\frac{2\alpha}{2+\alpha}, \;\; -1 < \beta \le 2, \ee
with the same range of $\alpha$ in view of (\ref{alfa}).

The orbit structure of the original family of potentials (\ref{pote}) at the energy level fixed by (\ref{ena}) will be approximated by the orbit structure of the rescaled Hamiltonian
\be\label{HaS}
\widetilde H = \frac12 \big(p_{x}^{2} + p_{y}^{2} \big) + \widetilde V_{\alpha}^{(N)} (x,y;q),\ee
where
\be\label{pots}
\widetilde V_{\alpha}^{(N)} (x,y;q) = L^{\beta} V_{\alpha}^{(N)} (x,y;L,q).\ee
The dynamics given by Hamiltonian (\ref{HaS}) take place in the rescaled time
\be\label{TS}
\tau = t/L^{\beta+1}\ee
at the new `energy'
\be\label{EL}
\widetilde E = L^{\beta} \left(E_{\alpha} - C_{(0,0)} (L,\alpha) \right) = 
  \frac1{\beta} \left(1-(L/L_{\rm max})^{\beta} \right).\ee
According to (\ref{amr}) the singular value $L=0$ is excluded from the analysis implying that the fictitious energy (\ref{EL}) is always finite and that the expansion around the equilibrium point (\ref{amin}) make sense.

The nonvanishing coefficients of the expansion of $\widetilde V_{\alpha}^{(N)}$ up to order $N=4$ are the following:
\ba
c_{(2,0)} = & \frac{2+\alpha}2, &\label{c20} \\
c_{(0,2)} = & \frac1{2 q^{2}}, &\label{c02} \\ 
c_{(3,0)} = & -\frac{10+3 \alpha-\alpha^{2}}6, &\label{c30} \\
c_{(1,2)} = & -\frac{2-\alpha}{2 q^{2}}, &\label{c12} \\
c_{(4,0)} = & -\frac{54+11 \alpha-6\alpha^{2}+\alpha^{3}}{24}, &\label{c40} \\
c_{(2,2)} = & -\frac{6-5 \alpha-\alpha^{2}}{4 q^{2}}, &\label{c22} \\
c_{(0,4)} = & -\frac{2-\alpha}{8 q^{4}}. &\label{c04} \ea
The first two of them provide the frequencies of the {\it epicyclic motions}. Recalling the time rescaling in (\ref{TS}), the radial and vertical harmonic frequencies are respectively
\be\label{rf}
\kappa = \frac{\sqrt{2+\alpha} }{ L^{\beta+1}} \ee
and
\be\label{vf}
\nu = \frac{1}{q L^{\beta+1}}. \ee

\section{Normal forms}

Hamiltonian (\ref{HaS}) is in the form of a power series and is therefore naturally apt to be treated in a perturbative way as a non-linear oscillator system. We construct a `normal form'  \citep{DB,SV}, namely a new Hamiltonian series which is an integrable approximation of the original one, suitably devised to catch its most relevant orbital features. 

The normal form is `non-resonant' when the two harmonic frequencies (\ref{rf}) and (\ref{vf}) are generically non-commensurable: in this case we get explicit formulas for actions and frequencies of the box orbits parented by the radial (disk) and by the vertical (thin tube) orbits. A `resonant' normal form is instead assembled by assuming from the start an integer value for the ratio of the harmonic frequencies and including in the new Hamiltonian terms depending on the corresponding resonant combination of the angles. This possibility might be considered an exception: it is instead the rule because, even if the unperturbed system is non-resonant with a certain real value 
\be \rho = \kappa/\nu \ee
of the frequency ratio, the non-linear coupling between the degrees of freedom induced by the perturbation, determines a `passage through resonance' with a commensurability ratio, say $m_1/m_2$, corresponding to the local ratio of oscillations in the two degrees of freedom. This in turn is responsible of the birth of new orbit families bifurcating from the normal modes or from lower-order resonances. 

To evaluate the most relevant resonances in our case, let us come back to the epicyclic frequencies (\ref{rf}) and (\ref{vf}). Their ratio is 
\be\label{RHO}
\rho = q \sqrt{2+\alpha}. \ee
We then approximate the frequencies with a rational number plus a small `detuning' \citep{CM,zm}
\be\label{DET}
\rho = \frac{m_1}{m_2} + \delta\ee
and proceed like as the unperturbed harmonic part would be in exact $m_1:m_2$ resonance putting the remaining part inside the `perturbation'. We speak of a {\it detuned} ($m_1$:$m_2$) {\it resonance}, with 
\be\label{Nm}
N_{\rm min} = m_1+m_2\ee 
the {\it order} of the resonance. In general, a normal form truncated at $N_{\rm min}$ includes the first resonant term.  With $\alpha$ in the range (\ref{alfa}) and an interval 
\be
0.5 < q < 1.5\ee
of reasonable values of $q$, we see that the most relevant resonance values with low commensurability are
\be
 \frac12, \; \frac23, \; \frac34, \; \frac11, \; \frac43, \; \frac32, \; \frac21.\ee
 In our analysis, we will mainly concentrate on the `central' value $1:1$ and on the higher-order cases $2:1$ and $4:3$. In fact, periodic orbits with these fequency ratios and associated quasi-periodic families are usually the most prominent in numerical investigations \citep{hu,conto2}. The procedure can be easily extended to other cases.
 
 Normal forms for the Hamiltonian system corresponding to (\ref{HaS}) are constructed with standard methods \citep{DB} and were used to determine the main features of the orbit structure of the logarithmic potential \citep{BBP2, PBB2}. We briefly resume the procedure in order to fix notations. After a scaling transformation
\ba
& x \longrightarrow (2+\alpha)^{1/4} \, x, \quad  
p_{x} \longrightarrow p_{x}/(2+\alpha)^{1/4}, \\
& y \longrightarrow y/\sqrt{q}, \quad  
p_{y} \longrightarrow \sqrt{q} p_{y},\ea
the original Hamiltonian 
 (\ref{HaS}) undergoes a canonical transformation to new variables $P_{X},P_{Y},X$ and $Y$, such that 
\begin{equation}\label{HK}
     K(P_{X},P_{Y},X,Y)=\sum_{n=0}^{N}K_n ,
  \end{equation}
generated by a function of the form 
\be\label{gene}
G=G_{1}+G_{2}+...\ee
with the prescription ($K$ in `normal form')
\be\label{NFD}
\{K_0,K\}=0.
\ee
In these and subsequent formulas, we adopt the convention of labeling the first term in the expansion with the index zero: in general, the `zero order' terms are quadratic homogeneous polynomials and terms of {\it order} {\it n} are polynomials of degree $n+2$. The zero order (unperturbed) Hamiltonian,  
\be\label{Hzero}
K_{0} \doteq \widetilde H_{0} = \frac12 \left(\omega_1 (P_X^2 + X^2) + \omega_2 (P_Y^2  +Y^2) \right),
\ee
with `unperturbed' frequencies 
\be \label{uf}
\omega_1=\kappa \ L^{1+\beta} = \sqrt{2+\alpha} ,  \quad
\omega_2 = \nu \ L^{1+\beta} = 1/q, \ee
plays, by means of the fundamental equation (\ref{NFD}), the double role of determining the specific form of the transformation and assuming the status of the second integral of motion. 
  The lowest order term of the generating function, $G_{1}$, is a cubic polynomial.
  
Using `action-angle'--{\it like} variables $\vec{J}, \vec{\theta}$ defined through the transformation
\ba
X &=& \sqrt{2 J_1} \cos \theta_1,\quad
Y  = \sqrt{2 J_2} \cos \theta_2,\label{AAV1}\\
P_X &=& \sqrt{2 J_1} \sin \theta_1,\quad
P_Y = \sqrt{2 J_2} \sin \theta_2,\label{AAV2}\ea 
the typical structure of the resonant normal form (truncated when the first resonant term appears) is \citep{SV, conto2}
\begin{eqnarray}\label{GNF}
K&=&m_{1} J_1+m_{2} J_2+ \sum_{k=2}^{m_1+m_2} {\cal P}^{(k)}(J_1,J_2)+\nonumber \\
&&a_{m_1 m_2} J_1^{m_{2}} J_2^{m_{1}} \cos [2(m_{2} \theta_{1}- m_{1} \theta_{2})], 
\end{eqnarray}
where ${\cal P}^{(k)}$ are homogeneous polynomials of degree $k$ whose coefficients may depend on $\delta$ and the constant $a_{m_1 m_2}(q,\alpha)$ is the only marker of the resonance. In these variables, the second integral is 
\be\label{cale}
{\cal E}=m_{1} J_1+m_{2} J_2\ee
and the angles appear only in the resonant combination
\be\label{psi}
\psi=m_{2} \theta_{1}- m_{1} \theta_{2}.\ee
For a given resonance, these two statements remain true for arbitrary $N>N_{\rm min}$. Introducing the variable conjugate to $\psi$,
\be\label{calr}
{\cal R}=m_{2} J_1-m_{1} J_2,\ee
the new Hamiltonian can be expressed in the {\it reduced} form $K({\cal R}, \psi; {\cal E},q,\alpha)$, that is a family of 1-dof systems parametrized by ${\cal E}$ (and $q,\alpha$). 

In the applications below we are interested in the global structure of phase-space, but the explicit solution of the equations of motion is also of great relevance. For a non-resonant normal form, the problem is easily solved: the coefficient $a_{m_1 m_2}$ vanishes and $K$ no longer has a term containing angles. Therefore, the $\vec{J}$ are `true' conserved actions and the solutions are
\be\label{SF}
X (\tau) = \sqrt{2 J_1} \cos \kappa_1 \tau ,\quad
Y (\tau) = \sqrt{2 J_2} \cos (\kappa_2 \tau + \theta_0),\ee
where
\be
\vec{\kappa} = \nabla_{\vec{J}} K\ee
is the frequency vector and $\theta_0$ is a suitable phase shift. 

In the resonant case instead, it is not possible to write the solutions in closed form. It is true that the dynamics described by the 1-dof Hamiltonian $K({\cal R}, \psi)$ are always integrable, but, in general, the solutions cannot be written in terms of elementary functions. However, solutions can still be written down in the case of the {\it main periodic orbits}, for which $\vec{J}, \vec{\theta}$ are true action-angle variables. There are two types of periodic orbits that can be easily identified:
\begin{enumerate}
      \item The {\it normal modes} for which one of the $\vec{J}$ vanishes.
         
      \item  The {\it periodic orbits in general position} characterized by a {\it fixed} relation between the two angles, $m_{2} \theta_{1}- m_{1} \theta_{2} \equiv \theta_0$.
       \end{enumerate}
In both cases, it is straightforward to check that the solutions retain a form analogous to Eq.(\ref{SF}) with known expressions of the actions and frequencies in terms of ${\cal E}$ and the parameters $ q$ and $\alpha$ such that $\kappa_1 / \kappa_2 = m_1 / m_2$. By using the generating function Eq.(\ref{gene}), the solutions in terms of standard `physical' coordinates can be recovered (apart from possible scaling factors) inverting the canonical transformation. As discussed in \citet{PBB2} the transformation back to the physical coordinates is expressed as a series of the form
\be
\vec{x} (\tau) = \vec{x}_{1} + \vec{x}_{2} + \vec{x}_{3} +...\ee   
and is given explicitly by
\ba
\vec{x}_{1} &=& \vec{X}, \label{x1}\\
\vec{x}_{2} &=& \{G_{1},\vec{X}\}, \label{x2} \\
\vec{x}_{3} &=& 
\{G_{2},\vec{X}\} + {\scriptstyle\frac12} \{G_{1},\{G_{1},\vec{X}\}\} \label{x3} \ea
and so forth. From a knowledge of the normalized solutions Eq.(\ref{SF}), we can therefore construct power series approximate solutions of the equations of motion of the original system
\be
\frac{d^{2} \vec{x}}{d \tau^{2}} = -\nabla_{\vec{x}} \widetilde V_{\alpha}^{(N)} (x,y;q).\ee

As a rule, normal modes exist on each `energy' surface $K=\widetilde E$. Periodic orbits in general position exist instead only beyond a certain threshold and we speak of a bifurcation ensuing from a detuned resonance. The bifurcation is usually described by a series expansion of the form
\be\label{detexp}
 \widetilde E = \sum_{k} a_{k} |\delta|^{k},\ee
where the $a_{k}$ are coefficients depending on the order $N_{\rm min}$ and the parameters $q,\alpha$. The order of truncation of the series is itself related to that of truncation of the normal form \citep{PBB1}. Eq.(\ref{detexp}) implies that at exact resonance (vanishing detuning) the bifurcation is intrinsic in the system and that, going away from the initial exact ratio of unperturbed frequencies, gradually increases the threshold value for the bifurcation.  We will see that already a linear relation given by the first order truncation provides a reliable estimate and will examine some example of a second order truncation.

\section{Applications}

We apply the theory resumed in the previous section to a set of typical bifurcation problems related to the potentials of section 2. In the following we adopt a `taxonomy' of the main resonant families that is coherent with our previous work and other standard references in the field \citep{BBP2,mes}. In case, we mention alternative denominations adopted by other authors.

\subsection{Non resonant box orbits}

We recall that we are studying the dynamics given by Hamiltonian (\ref{HaS}) 
at `energy' (\ref{EL}): a small value of it corresponds to a value of the angular momentum close to its maximum amount. This is the quasi harmonic regime in which the dynamics in the $x-y$ plane are characterized by {\it box} orbits oscillating around the two normal modes, $J_{1}=0$ and $J_{2}=0$. The first one is the $y$-axis periodic orbit and, coming back to the unscaled variables, it is the `vertical' $z$ oscillation: in 3 dimensions it gives the {\it thin tube} orbits. The second one is the $x$-axis periodic orbit that in the unscaled variables is the horizontal or `equatorial' $R$ oscillation and gives the {\it disk} orbits in 3 dimensions. 

A preliminary step is that of exploiting the non-resonant normal form. It can be readily used to get the post-epicyclic solution; more interestingly, we will exploit it later to attempt an approximate description of the higher-order resonances. The non-resonant normal form of the Hamiltonian up to the second order, takes the following form
 \be\label{KNR}
K  = \omega_1 J_1 +  \omega_2 J_2 + a J_1^{2} + b J_2^{2} + c J_1J_2,\ee
with
\ba
a &=& \frac{3}{4 \omega_{1}^2} \left(2 c_{(4,0)} \omega_{1}^2-5 c_{(3,0)}^2 \right), \label{knrag}\\
b &=& \frac{6 c_{(0,4)} \omega_{1}^2 (\omega_{1}^2 - 4 \omega_{2}^2) - 
          c_{(1,2)}^2 (3 \omega_{1}^2 - 8 \omega_{2}^2)}
          {4 \omega_{1}^2 \omega_{2}^2 (\omega_{1}^2 - 4 \omega_{2}^2)},\\
c &=& \frac{2 c_{(1,2)}^2 \omega_{1}^2 - (3 c_{(1,2)} c_{(3,0)} - c_{(2,2)} \omega_{1}^2) 
          (\omega_{1}^2 - 4 \omega_{2}^2)}{\omega_{1}^3 \omega_{2} (\omega_{1}^2 - 4 \omega_{2}^2)}.\label{knrcg}\ea
In the coefficients $b$ and $c$ it is appreciable the appearance of `small' denominators related to the 2:1 resonance, which is the first to appear in view of the lowest order coupling term $xy^2$. We also give the first term of the generating function (\ref{gene})
\ba \label{G1}
G_{1} &=& \frac{c_{(3,0)} P_X (2 P_X^2 + 3 X^2)}{\omega_{1}^{5/2}} +\\
           &&  \frac{c_{(1,2)}\left(P_X (\omega_{1}^2 Y^2 - 2 \omega_{2}^2 (P_Y^2  + Y^2)) \right)}
{\omega_{1}^{3/2} \omega_{2} (\omega_{1}^2 - 4 \omega_{2}^2)} - \\
&&  \frac{2 c_{(1,2)}  X Y P_Y}{\omega_{1}^{1/2} (\omega_{1}^2 - 4 \omega_{2}^2)}.\ea
Using expressions (\ref{c30}--\ref{c04}) and (\ref{uf}) the coefficients in the normal form can be written explicitly in terms of the parameters of the potential
\ba
a &=& \frac{-22 + 13 \alpha - \alpha^2}{24},\label{knra} \\
b &=& \frac{(2-\alpha) (10 + \alpha - 
   3 q^2 (2 +\alpha))}{4 q^2 (2 + \alpha) (-4 + q^2 (2 + \alpha))} , \label{knrb}\\
 c &=& \frac{(2 - \alpha) (6 - \alpha - q^2 (2 + \alpha))}{2 q \sqrt{
 2 + \alpha} (-4 + q^2 (2 + \alpha))}. \label{knrc}\ea
 The second order post-epicyclic solution is then given by the upgraded expression of the two frequencies of the 
 radial and vertical oscillations
\be\label{rf2}
\kappa = \frac1{ L^{\beta+1}}\left(\sqrt{2+\alpha} + 2 a J_1+ c J_2\right) \ee
and
\be\label{vf2}
\nu = \frac{1}{L^{\beta+1}} \left(\frac1{q}+2 b J_2 + c J_1\right) \ee
and by the orbit approximations (\ref{x1}--\ref{x2}) computed by means of the generating function (\ref{G1}). These results are the basis for attempting an accurate tracking of normal modes \citep{contos} and box orbits \citep{kent}.

\subsection{Bifurcations from the disk and from the thin tube orbit}
\label{bdtt}

We now start to illustrate the main body of results concerning the orbit structure as determined by the main bifurcations. 

Lowering the angular momentum (namely increasing the fictitious energy $\widetilde E$) both normal modes may lose their stability through a 1:1 resonance. We denote the bifurcating family as the {\it inclined} orbit in view of its natural interpretation as the {\it in phase} ($\psi=0$) 1:1 resonance of the two oscillations \citep{zm, BBP2}. The anti-phase ($\psi=\pi$) 1:1 resonant {\it loops} never appear in these systems, at least as a stable family (see subsection \ref{rtt}). The inclined periodic orbits parents two families of inclined boxes that may arrive quite far from the equatorial plane both above and below the disk: this phenomenon is called {\it levitation} \citep{ST1}. We recall that our inclined orbits have also been referred to as {\it reflected banana} by \citet{LS} and \citet{eva} and simply as {\it banana} by \citet{hu}: we prefer to leave this term as the standard denomination \citep{mes} for the 2:1 resonance.

To describe the bifurcation of the inclined orbit, the normal form is computed by a small `detuning' \citep{CM,zm} of the 1:1 resonance so that, from (\ref{DET}) with $m_1=m_2=1$,
\be\label{DET11}
\rho = q \sqrt{2+\alpha} = 1 + \delta. \ee
The normal form truncated to the first non zero resonant term is 
 \ba\label{K11}
\widetilde K &=& q K  = J_1 + J_2 +  \delta J_1 + q (a J_1^{2} + b J_2^{2} + c J_1J_2) \\
&& + d  J_1J_2 \cos [2(\theta_{1}- \theta_{2})],\ea
with $a,b,c$ as in (\ref{knrag}--\ref{knrcg}) and
\be\label{knrd}
d = \frac{q}{12} (1 + \alpha)(2 - \alpha).\ee 
The rescaling by $q$ lets to embody the detuned resonance (\ref{DET11}) in the neatest form. Introducing the resonant combinations
\be\label{psi11}
\psi=2 (\theta_{1}- \theta_{2})\ee
and
\be\label{calr11}
{\cal R}=J_1-J_2,\ee
the new Hamiltonian can be expressed in the {\it reduced} form 
\ba\label{KER11}
{\widetilde K} &=&  {\cal E} + 
\frac12 \delta ({\cal E} + {\cal R}) + A({\cal E}^{2} + {\cal R}^{2}) + B{\cal E}{\cal R} +\\
&& \frac14 ({\cal E}^{2} - {\cal R}^{2})  (qc + d \cos \psi), \ea
with 
\ba
A &=& \frac{q}4 (a+b)  ,\label{muA}\\
B &=& \frac{q}2 (a-b)  \label{muB}\ea
and 
\be\label{cale11}
{\cal E}=J_1+J_2\ee
is the second integral as in (\ref{cale}). Considering the dynamics at a fixed value of ${\cal E}$, we have that ${\widetilde K}$ defines a one--degree of freedom $(\psi,{\cal R})$ system with the following equations of motion
\ba
{\dot \psi} &=& {\widetilde K}_{\cal R} = 
\frac12 \delta + B{\cal E} + \frac12 \left(4A - (qc + d  \cos \psi)\right){\cal R},\label{dpsi}\\
{\dot {\cal R}} &=& - {\widetilde K}_{\psi} = 
\frac{d}4 \left({\cal E}^{2}  - {\cal R}^{2} \right) \sin \psi.\label{dr}
\ea
The fixed points of this system give the periodic orbits of the original system. The pair of fixed points with ${\cal R}=\pm {\cal E}$ correspond to the normal modes. The periodic orbits in ``general position'' (inclined and loops) are respectively given by the conditions $\psi = 0$ and $\psi = \pm \pi$ (which are the solution of ${\dot {\cal R}}=0$) and by the corresponding solutions of ${\dot \psi} = 0$:
\be\label{PO11}
{\cal R}=\frac{\delta + 2B{\cal E}}{qc-4A\pm d}.\ee
In view of (\ref{calr11}) and  (\ref{cale11}), the conditions
\be
0 \le J_1,J_2\le{\cal E},\ee
applied to the solutions (\ref{PO11}) translate into the conditions of existence
\be\label{EPO1}
{\cal E} \ge \frac{\delta}{q(c-2b) \pm d} \ee
and 
\be\label{EPO2}
{\cal E} \ge -\frac{\delta}{q(2a-c) \pm d}.\ee
The plus sign in front of $d$ corresponds to the loops and gives a case that will be examined in the following subsection. The conditions with the minus sign in front of $d$ correspond to the bifurcation of the inclined. Proceeding along the lines followed by \citet{BBP2}, it can be proven that, at the bifurcation, one of the normal modes suffers a stability--instability transition. The more common situation (for models ranging from sensibly oblate to prolate) is that in which the $x$-axis becomes unstable and the inclined appears as a pitchfork bifurcation from the disk orbit. The passage to instability of the $y$-axis is possible only for strongly oblate models and gives rise to a pitchfork  bifurcation from the thin tube.

\begin{table}
  \caption[]{Critical value of the angular momentum for the bifurcation of the inclined orbits from the disk orbit: the values in the third column are computed with Eq.(\ref{Lc11}) for general $\alpha$ and with eq.(\ref{LcL11}) for $\alpha=0$. The fourth and fifth columns list numerical predictions and their sources: \citet{LS} (LS); \citet{eva} (E); \citet{hu} (H); this work (T).}
  \begin{tabular}{@{}rrrrc@{}}
  \hline
      &            & \multicolumn{2}{c}{$L_{\rm crit}$}\\
$q$  & $\alpha$      & Analytical & Numerical & Source \\
\hline
0.4   & 0.5               & 0.27         & 0.33          & H          \\
0.5   & 0                  & 0.30         & 0.28          & T          \\
0.75 & 0                  & 0.42         & 0.419        & LS        \\
0.8   & 0                  & 0.28         & 0.26          & T          \\
0.8   & 0.1               & 0.24         & 0.22          & H          \\
0.85 & -0.18            & 0.23         & 0.29          & E          \\
\hline
\end{tabular}
\label{T1}
\end{table}

The bifurcation equations (\ref{EPO1},\ref{EPO2}) determine {\it critical} values of ${\cal E}$ in terms of the parameters $q,\alpha$. To make a quantitative prediction, we wont an expression for the corresponding {\it critical angular momentum}. The approach we have followed so far is altogether a perturbation approach truncated to the first non-trivial order. Therefore it is natural to look for expansions truncated to the first order in the detuning parameter. Taking into account the rescaling in $\widetilde K $ and the expressions  (\ref{knra}--\ref{knrc}) and (\ref{knrd}), the first order expansions of the critical values of the fictitious energy (\ref{EL}) are:
\be\label{c11}
\widetilde E = \cases { & $ \frac{12 (2+\alpha)}{5(-2-\alpha+\alpha^{2})}  \delta, \quad  \delta < 0,$\cr
                                    & $ \frac{6 (2+\alpha)}{2+\alpha-\alpha^{2}}  \delta, \quad  \delta > 0.$\cr}\ee
                                    The first solution corresponds to the bifurcation from the thin tube, the second one corresponds to the bifurcation from the disk. These are examples of series of the form (\ref{detexp}) truncated to the first order.
                                    
Using the relation between $\widetilde E$ and $L$ established by (\ref{EL}), we get the following expressions for the critical values of the angular momentum below which inclined orbits exist:
\be\label{Lc11}
L_{\rm crit} = \frac1{\sqrt{\rm e}} \times \cases { 
& $  \left(1- \frac{24 \alpha  (q \sqrt{2+\alpha} - 1)}{5(2+\alpha-\alpha^{2})} \right)^{-\frac{2+\alpha}{2\alpha}}, 
\quad  q < \frac1{\sqrt{2+\alpha}},$\cr
& $  \left(1+ \frac{12 \alpha (q \sqrt{2+\alpha} - 1)}{2+\alpha-\alpha^{2}} \right)^{-\frac{2+\alpha}{2\alpha}}, 
\quad  q > \frac1{\sqrt{2+\alpha}}.$\cr}\ee
It is also useful to write the limiting case of the logarithmic potential ($\alpha=0$):
\be\label{LcL11}
L_{\rm crit} = \cases { & $  {\rm e}^{-\frac{29}{10} + \frac{12}{5} \sqrt{2} q}, \quad  q < \frac1{\sqrt{2}},$\cr
                                   & $ {\rm e}^{\frac{11}2 - 6 \sqrt{2} q}, \quad  q > \frac1{\sqrt{2}}.$\cr}\ee

A comparison with the outcome of numerical determinations of the bifurcation threshold allows us to evaluate the accuracy of these analytical predictions. In Table \ref{T1} the critical value of the angular momentum for the bifurcation of the inclined orbits, computed with Eq.(\ref{Lc11}) for general $\alpha$ and with eq.(\ref{LcL11}) for $\alpha=0$, are compared with numerical data obtained either from published works \citep{LS,eva,hu} or by numerical computations made for the present paper. In this case, the bifurcation has been detected tracing the instability threshold of the normal mode by means of the Floquet method \citep{b3}. 

The accuracy is particularly good when the model is close to the exact resonance. Overall, the discrepancy linearly grows with detuning, as can be expected in this first order approach. The first two lines represent two strongly oblate models
with the thin tube to become unstable: in the rather extreme case with $q=0.4,\alpha=0.5$ the detuning is $\delta=-0.37$ and the relative error in the prediction is 18\%. In all other cases, the disk becomes unstable and the quality of the prediction can be represented by the case with $q=0.8,\alpha=0.1$ when the detuning is $\delta=0.16$ and the relative error in the prediction is 8\%. We may guess a prediction error
\be\label{PRED11}
\frac{L_{\rm crit,true}-L_{\rm crit}}{L_{\rm crit,true}} \simeq \frac12 \delta
= \frac12 (q \sqrt{2+\alpha} - 1)\ee
which can be used to further improve the accuracy of (\ref{Lc11}).

\subsection{Return to stability of the thin tube}
\label{rtt}

Pitchfork or period-doubling bifurcation are usually followed by a second stability change when the second resonant family appears \citep{mes}. In this setting, this possibility occurs if the loops appear. Using the bifurcation equations (\ref{EPO1},\ref{EPO2}) with the plus sign in front of $d$ and the explicit expressions of (\ref{knra}--\ref{knrc}), the inequality can be satisfied only with values of the parameters corresponding to rather extreme oblate models. This implies a negative value of the detuning and a critical fictitious energy
\be\label{cl11}
\widetilde E = -\frac{4 (2+\alpha)}{2+\alpha-\alpha^{2}}  \delta.\ee
We get the following expression for the critical value of the angular momentum below which loops bifurcate from the thin tube:
\be\label{Lcl11}
L_{\rm crit} = \frac1{\sqrt{\rm e}} \left(1- \frac{8 \alpha  (q \sqrt{2+\alpha} - 1)}{2+\alpha-\alpha^{2}} \right)^{-\frac{2+\alpha}{2\alpha}}. \ee                     
   This is again a pitchfork bifurcation so that the thin tube regains its stability. The loops are unstable and, lowering the angular momentum below the critical value, remain unstable for every reasonable combinations of the parameters. On the same ground, the inclined tend to occupy an even larger fraction of phase space. Referring again to the models in table \ref{T1}, in the case with $q=0.4,\alpha=0.5$ the critical value for the return to stability is $0.32$ \citep{eva} and (\ref{Lcl11}) predicts $0.18$.      

\subsection{Bifurcation of the inner thin tube}

\begin{table}
  \caption[]{Critical value of the angular momentum for the bifurcation of the inner tube orbits from the disk orbit: the values in the third column are computed with Eq.(\ref{Lc21}). The sources are as in the caption to Table \ref{T1}.}
  \begin{tabular}{@{}rrrrc@{}}
  \hline
      &            & \multicolumn{2}{c}{$L_{\rm crit}$}\\
$q$  & $\alpha$      & Analytical & Numerical & Source \\
\hline
1.1   & 0.25             & 0.51         & 0.41          & E          \\
1.1   & 0                  & 0.50         & 0.37          & T          \\
\hline
\end{tabular}
\label{T2}
\end{table}

A different phenomenon is that of the bifurcation of the {\it inner thin tube}. This happens only in prolate models and is due to a {\it period doubling bifurcation} from the equatorial orbit which becomes unstable at the bifurcation. In our terminology this is a {\it banana} orbit actually related to the 2:1 resonance. In the full three-dimensional problem, it gives rise to a thin tube that is always closer to the symmetry axis than the `outer' thin tube examined above which remains always stable \citep{dz}.

In this case, the detuning, from (\ref{RHO}) and (\ref{DET}) with $m_1=2, m_2=1$, is
\be\label{DET21}
\delta = q \sqrt{2+\alpha} - 2. \ee
The normal form truncated to the first non zero resonant term is 
 \be\label{K21}
\widetilde K  = 2 J_1 + J_2 +  \delta J_1 + D  \sqrt{J_1} J_2 \cos (\theta_{1}- 2\theta_{2})\ee 
with
\be
D=\frac{q^{2} c_{(1,2)}}{\sqrt{2} (2+\alpha)^{1/4}}=
\frac{\alpha - 2}{2 \sqrt{2} (2+\alpha)^{1/4}}.\ee
Introducing the resonant combinations
\be\label{psi21}
\psi=\theta_{1}- 2\theta_{2}\ee
and
\be\label{calr21}
{\cal R}=J_1-2J_2,\ee
the new Hamiltonian can be expressed in the reduced form 
\be\label{KER21}
{\widetilde K} =  {\cal E} + 
\frac15 \delta (2{\cal E} + {\cal R}) + \frac{D}{\sqrt{5} \ 5} \sqrt{2{\cal E}+ {\cal R}} ({\cal E} - 2{\cal R}) \cos \psi, \ee
where now
\be\label{cale21}
{\cal E}=2J_1+J_2.\ee

The fixed points corresponding to the periodic orbits in general position are given by the solutions of
\ba
 D (7{\cal E}+6 {\cal R}) - 2 \sqrt{5} \sqrt{2{\cal E}+ {\cal R}} \delta &=& 0, \;\; \psi = 0, \\
 D (7{\cal E}+6 {\cal R}) + 2 \sqrt{5} \sqrt{2{\cal E}+ {\cal R}} \delta &=& 0, \;\; \psi = \pi.\ea
 Combining these with the existence conditions 
 \be
0 \le J_1\le{\cal E}/2, \; \; 0 \le J_2\le{\cal E},\ee
gives the critical value
\be
{\cal E} = \frac1{2 D^{2}} \delta^{2}\ee
and, taking into account the rescaling implicit in the normal form,
\be\label{c21}
\widetilde E = \frac{2 (2+\alpha)}{(2-\alpha)^{2}} \delta^{2}.\ee
We now have an example of a series of the form (\ref{detexp}) in which the first coefficient vanishes and is therefore truncated to the second order. Using  (\ref{EL}) the prediction of the value of the angular momentum below which the inner thin tube bifurcates from the disk is
\be\label{Lc21}
L_{\rm crit} = \frac1{\sqrt{\rm e}} \left(1+ \frac{4 \alpha  (q \sqrt{2+\alpha} - 2)^{2}}{(2-\alpha)^{2}} \right)^{-\frac{2+\alpha}{2\alpha}}. \ee
The pure logarithmic $\alpha=0$ limit is given by
\be\label{Lcl21}
L_{\rm crit} = {\rm e}^{-\frac{11}2 + 4 \sqrt{2} q - 2 q^{2}}. \ee
In Table \ref{T2} the critical value of the angular momentum for the bifurcation of the inner thin tube orbits, computed with Eq.(\ref{Lc21}), are compared with numerical data. We may guess a prediction error
\be\label{PRED21}
\frac{L_{\rm crit,true}-L_{\rm crit}}{L_{\rm crit,true}} \simeq 2 \delta^{2}
= 2 (q \sqrt{2+\alpha} - 2)^{2}\ee
which can be used to further improve the accuracy of (\ref{Lc21}).

\subsection{Bifurcations of higher-order resonant orbits}

As an application of the approach adopted by \citet{SV} we compute the appearance of the {\it pretzel} as a very high-order (4:3) resonance. In the literarure it is also denoted as a `reflected' fish \citep{LS,eva}. In spite of the high order, this family happens to pay a relevant role in shaping the phase space of these systems, occupying a substantial fraction of surfaces of section for a wide range of parameters in oblate models \citep{LS,eva,hu}.

The implementation of the general procedure followed in all cases treated above would require, in the present instance, a normal form truncated to $N_{\rm min}=7$ in order to include at least the first resonant terms. The algebraic manipulators available nowadays have no problem in accomplishing this endeavour without excessive CPU times. However, this implies a huge number of terms which hinders a completely general algebraic approach. The main troubles come from the solution of the equations for the fixed points of the reduced system, giving in turn the periodic orbits in general position. Being polynomials of degree one less than the normal form in the action variables, these give rise to algebraic equations of which it is very difficult to write the solutions in a manageable way. A straighforward approach would be that of examining specific cases and numerically solving for their roots. However, this spoils this general method of all its appeal. An alternative approach is to give up for a detailed knowledge of the periodic orbits in general position and try an approximate location of their `resonance manifold' together with its first order condition of existence \citep{SV}.

We recall that, for a general reduced Hamiltonian of the form $K({\cal R}, \psi; {\cal E},q,\alpha)$, the fixed points with $\psi = 0$ and  $\psi = \pm\pi$ correspond to two different orbit sets with two given values of ${\cal R}(0)$ and ${\cal R}(\pi)$. They can be separately identified only if the normal form contains at least the first resonant term. The method adopted by \citet{SV}, by truncating at a lower order term, only allows us to find an `average' value of the two ${\cal R}$ coordinates together with an existence condition, a linear or quadratic equation in the simplest instances. However, it involves the detuning and the second approximate integral and therefore provides an estimate of the critical energy for the bifurcation. 

In this case the detuning, from (\ref{RHO}) and (\ref{DET}) with $m_1=4, m_2=3$, is
\be\label{DET43}
\delta = q \sqrt{2+\alpha} - \frac43. \ee
The truncated normal form is 
 \be\label{K43}
\widetilde K  = 4 J_1 + 3 J_2 + 3 \delta J_1 + 3q(a J_1^{2} + b J_2^{2} + c J_1J_2)+...\ee 
where the second order term is essentially the same as in the non-resonant normal form (\ref{KNR}) and the dots stand as a remainder of the possible necessity of continuing the expansion in case the quality of the approximation is too low. 
In this case one can simply chose
\be\label{calr43}
{\cal R}=-J_1\ee
and
\be\label{cale43}
{\cal E}=4J_1+3J_2.\ee
The non constant terms of the reduced Hamiltonian are
\be\label{KER43}
- 3 \delta  {\cal R} + q\left[\left(a+\frac{16b}3 - 4c\right){\cal R}^{2}+\left(\frac{8b}3 - c\right){\cal E}{\cal R}\right]+... \ee
The resonance manifold is determined by the condition $\dot \psi = \widetilde K_{\cal R} = 0$ so that we have 
\be\label{R43}
{\cal R}_{RM}=\frac{ 27 \delta-8 b {\cal E} q + 3 c {\cal E} q}{2 (9 a + 16 b - 12 c) q}.\ee
Combining this with the existence condition
\be
-\frac14  {\cal E} \le {\cal R}_{RM} \le 0,\ee
the rescaling $\widetilde E = 3 q {\cal E} $ and the expressions (\ref{knra}--\ref{knrc}), we get
the following second order approximation for the critical `energy'
\be\label{c43}
{\widetilde E}_{RM} = -\frac{15}{\sqrt{2(2+\alpha)}} \delta - \frac{21}{2+\alpha} \delta^{2}.\ee
Using  (\ref{EL}), the prediction of the value of the angular momentum below which the inner thin tube bifurcates from the disk is
\be\label{Lc43}
L_{\rm crit} = \frac1{\sqrt{\rm e}} \left(1- \beta {\widetilde E}_{RM} (q,\alpha) \right)^{-\frac{2+\alpha}{2\alpha}}. \ee

In Table \ref{T3} the critical value of the angular momentum for the bifurcation of the inclined orbits, computed with Eq.(\ref{Lc43}), are compared with numerical data. The accuracy reached is unexpectedly good in view of the rough nature of the approximating technique. This method could also be employed to investigate the properties of resonant boxlets of triaxial ellipsoids \citep{Zhao,ZZ}.

\begin{table}
  \caption[]{Critical value of the angular momentum for the bifurcation of the 4:3 pretzel (reflected fish). The sources are as in the caption to Table \ref{T1}.}
  \begin{tabular}{@{}rrrrc@{}}
  \hline
      &            & \multicolumn{2}{c}{$L_{\rm crit}$}\\
$q$  & $\alpha$      & Analytical & Numerical & Source \\
\hline
0.75 & 0                  & 0.17         & 0.173          & LS        \\
0.8   & -0.3              & 0.10           & 0.15          & H          \\
0.85 & -0.18            & 0.18           & 0.16          & E          \\
0.9   & 0                  & 0.40           & 0.38          & T          \\
\hline
\end{tabular}
\label{T3}
\end{table}

\section{Discussion and conclusions}

Motion in the meridional plane of an axisymmetric system is a stumbling block and a case study in dynamical astronomy. Its investigation has a long history and has inspired the development of many ideas and methods \citep{conto1,hh, hori,gu, fv}. 
The development of techniques to investigate the regular and chaotic structure of phase-space and of approximating integrals of motion (a `third integral' in this case, in addition to $E$ and $L$) pay much to the activity on this topic. In particular, a systematic investigation of the regular dynamics by constructing Hamiltonian `normal forms' started with the work of \citet{hori} and \citet{gu} and, suitably traduced in algorithms of normalization, is at the basis of many subsequent works including the present one.

Here we have shown how a simple first-order truncation of the normal form is able not only to convey qualitative information on the phase-space structure but also quantitative predictions with concrete meaning for applications in galactic dynamics. In specific situations, especially in non-scalefree models, painstaking extended numerical simulations are usually required to even get a flavor of what is going on. A simple albeit rough analytical recipe can therefore be quite useful to get an average picture. With this approach, in place of numerical simulations quite heavy in view of the large parameter space, simple analytical formulas are presented for the computation of the main bifurcation thresholds of the problem. For each scalefree potential with slope given by $\alpha$ and ellipticity $q$, a critical value of the angular momentum of the full tridimensional system determines the appearance of some new orbit family and, possibly, a change in the stability nature of the `parent' orbit. The approach can clearly be generalized to non-scalefree models at the price of a fourth parameter in addition to $L,\alpha$ and $q$. 

Overall, these results make us confident that, in order to improve the accuracy or to explore the features of minor families and/or higher-order resonances, it can be worth the effort of constructing and investigating complete resonant normal forms. However, it is important to remark that the asymptotic nature of the results implies a trade-off between accuracy and extent of the phase-space region that is mimicked by the normal form. This region is largest at some {\it optimal truncation order} \citep{ECG} that depends on the nature of the system and the order of the resonance. The accuracy in the predictions of e.g. the bifurcation thresholds can be further improved with respect to that at the optimal order but only at the expense of reconstruction of dynamics. The choice of the whole strategy is then determined by the needs of the problem at hands.

\section*{Acknowledgments}

The author acknowledges the financial support of INFN (Sezione di Roma Tor Vergata) and of Icranet (International center for Relativistic Astrophysics Network).

\label{lastpage}

\end{document}